%% file: main.tex
\documentclass[10pt,letterpaper,twocolumn]{extarticle}

\usepackage[
  letterpaper,
  top=0.75in, bottom=1.0in,
  left=0.75in, right=0.75in,
  columnsep=0.33in
]{geometry}

\usepackage[T1]{fontenc}
\usepackage[utf8]{inputenc}
\usepackage{libertine}                         % Linux Libertine body text
\usepackage[scaled=0.97]{inconsolata}          % Inconsolata for \texttt
\usepackage[libertine,vvarbb]{newtxmath}       % Math compatible with Libertine
\usepackage{textcomp}
\usepackage{microtype}

\usepackage{amsmath,mathtools}

\usepackage{graphicx}
\usepackage{booktabs}
\usepackage{array}
\usepackage{multirow}
\usepackage{tabularx}
\usepackage{caption}
\usepackage{subcaption}
\usepackage{enumitem}
\usepackage{float}

\usepackage{titlesec}
\titleformat{\section}
  {\normalfont\large\bfseries\MakeUppercase}{\thesection}{0.6em}{}
\titleformat{\subsection}
  {\normalfont\normalsize\bfseries}{\thesubsection}{0.6em}{}
\titleformat{\subsubsection}
  {\normalfont\normalsize\itshape}{\thesubsubsection}{0.6em}{}
\titlespacing*{\section}      {0pt}{1.4ex plus .5ex minus .2ex}{0.7ex plus .2ex}
\titlespacing*{\subsection}   {0pt}{1.1ex plus .4ex minus .2ex}{0.5ex plus .2ex}
\titlespacing*{\subsubsection}{0pt}{0.9ex plus .3ex minus .2ex}{0.4ex plus .2ex}

\usepackage{fancyhdr}

\usepackage{balance}

\usepackage{xcolor}
\usepackage[hidelinks,breaklinks=true]{hyperref}
\usepackage{url}
\usepackage[numbers,square,sort&compress]{natbib}

\usepackage{comment}
\excludecomment{CCSXML}
\newcommand{\ccsdesc}[2][]{}
\providecommand{\Description}[2][]{}
\providecommand{\authornote}[1]{}
\providecommand{\authorsaddresses}[1]{}
\providecommand{\setcopyright}[1]{}
\providecommand{\copyrightyear}[1]{}
\providecommand{\acmYear}[1]{}
\providecommand{\acmConference}[4][]{}
\providecommand{\acmBooktitle}[1]{}
\providecommand{\acmISBN}[1]{}
\providecommand{\acmDOI}[1]{}
\providecommand{\acmPrice}[1]{}
\providecommand{\settopmatter}[1]{}
\providecommand{\teaserfigure}[1]{}
\providecommand{\received}[2][]{}
\providecommand{\editor}[1]{}

\newenvironment{acks}{\section*{Acknowledgments}}{}

\makeatletter

\let\@orig@title\title
\renewcommand{\title}[2][]{\@orig@title{#2}}

\newcommand{\@subtitletext}{}
\newcommand{\subtitle}[1]{\gdef\@subtitletext{#1}}

\newcommand{\@titlethanks}{}
\renewcommand{\thanks}[1]{\gdef\@titlethanks{#1}}

\newcommand{\institution}[1]{#1}
\newcommand{\city}[1]{\unskip, #1}
\newcommand{\country}[1]{\unskip, #1}

\newcommand{\@authorblocks}{}
\newcommand{\@curauthor}{}
\newcommand{\@curaffil}{}
\newcommand{\@curemail}{}
\newif\if@haveauth \@haveauthfalse

\newcommand{\@oneauthor}[3]{%
  \begin{minipage}[t]{0.45\textwidth}%
    \centering
    {\bfseries #1}\\[2pt]%
    #2\\[2pt]%
    \texttt{#3}%
  \end{minipage}\hspace*{\fill}%
}

\newcommand{\@flushauthor}{%
  \if@haveauth
    \protected@edef\@temp{%
      \noexpand\@oneauthor
        {\unexpanded\expandafter{\@curauthor}}%
        {\unexpanded\expandafter{\@curaffil}}%
        {\unexpanded\expandafter{\@curemail}}%
    }%
    \expandafter\g@addto@macro\expandafter\@authorblocks\expandafter{\@temp}%
  \fi
  \global\let\@curauthor\@empty
  \global\let\@curaffil\@empty
  \global\let\@curemail\@empty
}

\renewcommand{\author}[1]{%
  \@flushauthor
  \gdef\@curauthor{#1}%
  \global\@haveauthtrue
}
\newcommand{\affiliation}[1]{\gdef\@curaffil{#1}}
\providecommand{\email}{}\renewcommand{\email}[1]{\gdef\@curemail{#1}}

\newcommand{\keywords}[1]{%
  \par\noindent\textbf{Keywords:} #1\par\medskip
}

\renewcommand{\@maketitle}{%
  \@flushauthor
  \newpage
  \null
  \begin{center}%
    {\LARGE\bfseries \@title \par}%
    \vskip 0.5em
    \ifx\@subtitletext\@empty\else
      {\large\itshape \@subtitletext
        \ifx\@titlethanks\@empty\else
          \protect\footnotemark[1]%
        \fi\par}%
    \fi
    \vskip 1.4em
  \end{center}%
  \noindent\hspace*{\fill}\@authorblocks\par
  \vskip 1.2em
}

\renewcommand{\maketitle}{%
  \par
  \begingroup
    \renewcommand{\thefootnote}{\fnsymbol{footnote}}%
    \if@twocolumn
      \twocolumn[\@maketitle]%
    \else
      \@maketitle
    \fi
    \ifx\@titlethanks\@empty\else
      \footnotetext[1]{\@titlethanks}%
    \fi
  \endgroup
  \setcounter{footnote}{0}%
}

\makeatother

\title[Bridging Explainable AI and Human Agent Teaming]{Evaluating XAI Support From A Hierarchical Reinforcement Learning Policy in Human-Agent Collaboration}
\subtitle{Full version of paper accepted as an Extended Abstract at AAMAS 2026}
\thanks{Supplementary materials (Appendices A and B), the AAMAS 2026 poster, and the Extended Abstract are available at \url{https://github.com/mateuslevisf/xai-hat-aamas2026}.}

\author{Mateus Levi Simões Fernandes}
\affiliation{%
  \institution{PUC-Rio}%
  \city{Rio de Janeiro}%
  \country{Brazil}}
\email{mfernandes@inf.puc-rio.br}

\author{Alberto Sardinha}
\affiliation{%
  \institution{PUC-Rio}%
  \city{Rio de Janeiro}%
  \country{Brazil}}
\email{sardinha@inf.puc-rio.br}

\begin{document}

\pagestyle{fancy}
\fancyhead{}
\renewcommand{\headrulewidth}{0pt}

\maketitle

%%% Abstract
\begin{abstract}
Explainable AI (XAI) has shown promise for human-agent collaboration, yet results rely on hand-crafted policies in custom environments, limiting generalizability to state-of-the-art teaming research. We provide the first systematic evaluation of XAI support generated from an intrinsically explainable learned policy in an established benchmark. Using the Hierarchical Ad Hoc Agents (HA$^2$) architecture in Overcooked-AI, we generate real-time explanations from hierarchical subtask selections, delivered through text or audio via a novel trigger-based system. Our between-subjects experiment (n=38) found no significant performance effects, though participants with explanations showed trends toward faster performance improvement. More notably, audio explanations produced a significant reduction in participants' working-alliance bond with the agent -- an effect absent under the text modality -- suggesting that spoken explanations activate partnership expectations the underlying reactive policy cannot meet. We provide the first modality comparison in real-time human-agent collaboration and establish a baseline methodology for evaluating intrinsically explainable reinforcement learning architectures in benchmark environments. Results point to matching explanation modality to the underlying policy's capacity of sustaining the partnership its delivery implies as a potential path for more effective collaborative XAI.
\end{abstract}



%%%%%%%%%%%%%%%%%%%%%%%%%%%%%%%%%%%%%%%%%%%%%%%%%%%%%%%%%%%%%%%%%%%%%%%%

\input{chapters/introduction}
\input{chapters/related-work}
\input{chapters/methodology}
\input{chapters/results}
\input{chapters/conclusions}

%%%%%%%%%%%%%%%%%%%%%%%%%%%%%%%%%%%%%%%%%%%%%%%%%%%%%%%%%%%%%%%%%%%%%%%%

\section*{Data and Materials Availability}
Supplementary materials accompanying this paper are available at \url{https://github.com/mateuslevisf/xai-hat-aamas2026}, including (1) the complete trigger-based explanation templates with variable mappings and implementation details, (2) full survey instruments for all measures used in the study, (3) the AAMAS 2026 poster, and (4) the published Extended Abstract.

\begin{acks}
This study was approved by PUC-Rio's Research Ethics Committee (Protocol SGOC 545928, approved June 11, 2025). This study was financed in part by the Coordenação de Aperfeiçoamento de Pessoal de Nível Superior - Brasil (CAPES) - Finance Code 001.

This paper was accepted and published as an Extended Abstract at AAMAS 2026.
\end{acks}

\appendix

\section{Human-Agent Fluency Questionnaire: Adaptations from Hoffman (2019)}
\label{app:fluency-adaptation}
The deployed instrument adapts the Human-Robot Fluency Assessment \cite{hoffman2019evaluating} along four dimensions:

\begin{enumerate}

      \item \textbf{Terminology.} All occurrences of the word ``robot'' were replaced with ``agent'' throughout the questionnaire to reflect the software-agent nature of the system evaluated in this study. The instrument was deployed bilingually (English and Brazilian Portuguese), with anchor direction preserved across both translations.

      \item \textbf{Item insertion.} A non-Hoffman item was inserted as Q5, added by the experimenters as a sharper rephrasing of Q4: \emph{``I had to take charge to ensure the human-agent team performed well.''} This insertion shifts the canonical Hoffman Q-numbers by $+1$ for items falling between Q5 and Q11 in the deployed numbering (deployed Q12 onward re-aligns with Hoffman's numbering).

      \item \textbf{Expanded Robot Relative Contribution subscale.} Hoffman's 4-item Relative Contribution subscale (Q4--Q7 in his numbering) was expanded to 5 items in our deployment (Q4--Q8), incorporating the experimenter-added Q5 alongside the four Hoffman items.

      \item \textbf{Reverse-coded item set.} Reverse-coding was applied to $\{$Q4, Q5, Q7, Q16, Q24$\}$ on the deployed instrument, each scored as $8 - x$ on the 1--7 scale. Relative to Hoffman's reverse-coded set mapped onto the deployed numbering ($\{$Q4, Q7, Q16, Q24$\}$), the only addition is the experimenter-authored Q5, which parallels Q4 in direction.

  \end{enumerate}

All other subscale structures, item orderings, and response scales follow Hoffman's \citeyear{hoffman2019evaluating} published instrument. Item Q10(``The agent was trustworthy.'') appears in both the Trust in Robot and Positive Teammate Traits subscales; this dual-use compensates for Hoffman's instrument including a duplicate ``trustworthy'' item (his Q9 and Q11) that the deployed survey rendered only once.

\bibliographystyle{plainnat}
\bibliography{main}

\end{document}

%% file: chapters/introduction.tex
\section{Introduction}

As artificial intelligence systems transition from tools to collaborative partners, they must become not only effective in executing tasks but also comprehensible to their human teammates. In human-human collaboration, teammates communicate intentions and negotiate strategies using shared mental models. But human-AI collaboration often lacks mutual understanding due to agents having opaque decision-making processes, frequently based on hard-to-comprehend neural network architectures. This understanding gap contradicts McDermott's observation that systems must be both correct \textit{and} understood \cite{mcdermott1978tarskian}, creating a fundamental challenge: if AI agents are to occupy roles traditionally reserved for humans, they must operate at levels of transparency similar to human teammates.

Explainable artificial intelligence (XAI) research offers promising approaches to bridge this gap \cite{sado2023explainable}, yet systematic evaluation of such techniques in collaborative contexts remains limited. Most human-agent teaming research prioritizes optimizing agent performance over leveraging human understanding \cite{strouse_collaborating_2022,loo_hierarchical_2023}, treating humans as unpredictable variables rather than partners whose comprehension of the autonomous agent could enhance coordination. While Paleja et al.~\cite{paleja2021utility} demonstrated that XAI-based support can improve collaboration, their work relied on hand-crafted policies in a custom Minecraft environment. This limits transferability of their conclusions to state-of-the-art human-agent teaming research, which uses established benchmarks like Overcooked-AI \cite{carroll_utility_2020}. Such a context reveals a critical methodological gap: no work has systematically evaluated XAI support generated from learned, high-performing policies in benchmark collaborative environments.

We address this gap by leveraging the Hierarchical Ad Hoc Agents (HA$^2$) \cite{arocaouellette2025implicitlyaligninghumansautonomous} architecture, which achieves state-of-the-art performance when collaborating with humans in Overcooked-AI, through policies trained on a human-interpretable subtask decomposition. HA$^2$'s hierarchical structure displays intrinsic interpretability \cite{vilone2020explainable}: any action taken by the agent on the environment depends on comprehensible high-level subtask choices. Even so, this explanatory potential has not been systematically evaluated by the authors or succeeding work. To remedy this, we generate real-time explanations directly from the agent's hierarchical decision-making process and deliver them through text or audio modalities via a novel trigger-based system.

Our between-subjects experiment (n=38) examines both immediate effects and adaptation patterns across four collaborative sessions, investigating whether explanations improve objective performance and subjective metrics, and whether audio delivery reduces cognitive interference in visually demanding tasks. We find that simple status explanations provide no immediate performance benefit, though participants receiving audio explanations show trends toward faster improvement rates, suggesting XAI may enhance adaptation rather than immediate task execution. The subjective-experience analysis, however, identifies a Bonferroni-significant cost specific to spoken delivery: audio explanations reduced participants' working-alliance bond with the agent in a way that text and the absence of explanations did not, a pattern we interpret as evidence that spoken delivery activates partnership expectations the underlying reactive policy cannot meet. Our sample's high mean gaming familiarity (identified through post-hoc analysis) provides additional context for interpreting the performance-related nulls.

This work makes three key contributions: (1) the first systematic XAI evaluation using state-of-the-art learned policies in an established teaming benchmark, demonstrating that intrinsically explainable reinforcement learning architectures can generate authentic explanations for real-time collaboration; (2) the first comparison of explanation modality effects (text vs. audio) in a real-time human-agent collaborative task; and (3) a baseline trigger-based delivery system for hierarchical policy explanations. These insights advance understanding of when and how XAI operates in fast-paced collaborative environments, providing guidance for designing explanation systems that match not only user expertise and task demands but also the underlying policy's capacity to sustain the partnership its delivery implies.

%% file: chapters/related-work.tex
\section{Related Work}

The challenge of making AI agents comprehensible during collaboration sits at the intersection of two research communities with complementary blindspots. Explainable AI research has developed sophisticated techniques for transparency but struggles with systematic evaluation in realistic collaborative contexts. Human-agent teaming research has created benchmark environments and state-of-the-art learning algorithms but largely ignores the social dimensions of collaboration, treating agent transparency as secondary to performance optimization.

\textbf{XAI in Human-Agent Collaboration:} One of the most comprehensive evaluations of XAI utility during real-time human-agent collaboration is found in Paleja et al.'s using Minecraft \cite{paleja2021utility}, which demonstrated that the effectiveness of XAI support depends critically on user expertise and the type of explanation given by the agent. Novice participants showed improved performance when receiving status-based explanations, but not with full decision tree visualizations. Meanwhile, expert participants experienced performance degradation when receiving explanations, presumably due to higher cognitive load disrupting fluid coordination. Subjective measures told a different story: participants consistently rated explainable agents more positively regardless of performance effects, suggesting explanations can still provide subjective benefits beyond immediate task optimization.

Paleja's work established important principles: that real-time explanations differ fundamentally from post-hoc analysis; that modality and form of the explanations matter for fast-paced tasks; and that task expertise moderates explanation utility. Yet two critical limitations constrain its applicability to modern human-agent teaming research. First, the agent used a hand-crafted hierarchical policy rather than learned behaviors, making it unclear whether their findings transfer to deployable machine learning systems. Hand-crafted policies enable perfect control over explanation content and timing but sidestep the fundamental challenges of generating explanations from learned decision-making processes. Second, the work was based on a custom Minecraft environment where the agent had limited capabilities -- serving rather as an assistant gathering or constructing resources for the human participant -- instead of being an equal partner. This limits generalizability when we consider established collaborative benchmarks, where coordination faces additional challenges and (or because) teammates have equal capabilities.

Recent work by Wang et al. \cite{wang2024utility} addresses some of these limitations by developing external Theory of Mind models that predict future actions of black-box agents in the benchmark Overcooked-AI environment. By training transformer-based predictors on offline agent trajectories, their approach generates visual predictions of upcoming primitive actions without requiring the agent itself to be interpretable. While this demonstrates that post-hoc explanation systems can improve human-AI coordination even for opaque agents, their approach relies on substantial offline data collection and separate model training pipelines -- factors that are avoidable if a model's intrinsic explainability can be leveraged. The visual prediction modality they employ also differs from simpler explanation approaches based on natural language, leaving open questions about how different modality and content choices would affect explanation utility.

Prior work exploring communication in collaborative contexts reinforces both the promise and challenges of agent transparency. Early work by Harbers et al. \cite{harbers_explanation_2011} found that agent explanations improved subjective experience without enhancing performance, attributing this to cognitive overhead. Mercado et al. \cite{mercado2016intelligent} showed that transparent AI assistants increased both trust and performance in unmanned vehicle operations, though using domain-specific transparency models. More recently, Zhang et al. \cite{zhang_investigating_2023} demonstrated through a Wizard-of-Oz study that humans prefer proactive AI communication while noting that excessive communication creates distraction, particularly after coordination patterns stabilize. These findings collectively suggest that explanation utility depends on matching communication frequency and content to both task demands and human expertise levels, but none of these works evaluated explanations generated from state-of-the-art learned policies in a benchmark teaming environment.

\textbf{Human-Agent Collaboration in Overcooked-AI:} Carroll et al. \cite{carroll_utility_2020} established Overcooked-AI as the \textit{de facto} benchmark for human-agent collaboration research, demonstrating that agents trained exclusively with other AI partners fail catastrophically when teaming with humans. Their Behavior-Cloning Play (BCP) method addressed this by training agents on human gameplay data, but subsequent work challenged the necessity of human data entirely. Strouse et al.'s Fictitious Co-Play (FCP) \cite{strouse_collaborating_2022} achieved superior performance by training agents as best responses to diverse self-play populations rather than specific human behaviors, establishing that exposure to varied teammate profiles enables effective human collaboration without the necessity of data generated from human behavior.

This insight that robustness to diverse teammates enables human collaboration motivated increasingly sophisticated population-based training approaches. Loo et al.'s Hierarchical Population Training (HiPT) \cite{loo_hierarchical_2023} extended FCP by learning multiple low-level policies alongside a high-level manager that dynamically selects the appropriate policy which serves as the best response for the current teammate. This meta-learning approach, which trains the agent to recognize and adapt to different teammate types, achieved state-of-the-art performance by making the agent robust not just to diverse behaviors but to varying skill levels.

The Hierarchical Ad Hoc Agents (HA$^2$) architecture \cite{arocaouellette2025implicitlyaligninghumansautonomous} represents a conceptual shift from pure performance optimization toward human-aligned task representations. Rather than learning multiple best response policies like HiPT, HA$^2$ provides agents with the same hierarchical task structures humans naturally use for coordination, decomposing complex collaborative tasks into mutually understandable subtasks and using this hierarchy as basis for the agent's decision-making. This approach leverages the insight that effective human collaboration emerges not just from adapting to diverse behaviors but from operating within shared cognitive frameworks \cite{sebanz2006joint,tenenbaum2011grow}.

Critically for our purposes, HA$^2$ achieves this human alignment while maintaining state-of-the-art performance in Overcooked-AI through a Manager-Worker architecture. The Manager selects high-level subtasks (e.g., ``pick up onion from dispenser'', ``place onion in pot'') while a Worker policy executes the  low-level actions to complete the chosen subtask. The authors argue this hierarchical structure inherently supports transparency: any action the agent takes can be traced to a human-interpretable subtask choice. Yet this claim of interpretability has never been empirically validated, since the original work focused exclusively on implicit coordination through shared task abstractions without evaluating whether explicitly communicating these subtask choices actually improves human-agent collaboration.

This creates our core opportunity: HA$^2$ provides both high performance and intrinsic explainability through its hierarchical decomposition. Unlike hand-crafted policies \cite{paleja2021utility} that sacrifice deployability for interpretability or opaque deep learning policies that require external models and separate training \cite{wang2024utility}, HA$^2$'s architecture naturally generates human-interpretable decision representations as part of its normal operation. The Manager's subtask selections occur at a level of temporal and semantic abstraction that matches how humans naturally describe collaborative intentions: "I'm getting an onion" rather than "moving south three tiles then pressing interact."

By exposing these subtask choices through real-time explanations, we can evaluate XAI utility using a state-of-the-art learned policy in an established benchmark environment. Furthermore, we systematically compare text versus audio delivery to understand how modality choices affect explanation utility in visually demanding tasks -- something prior work has not evaluated despite using different modalities \cite{paleja2021utility, wang2024utility}. This methodological advance enables more rigorous investigation of when and how XAI operates in realistic human-agent teaming contexts.

This review reveals three unexplored opportunities our work addresses. First, no systematic XAI evaluation exists in established HAT benchmarks. While Paleja et al. demonstrated promising results, their custom environment and hand-crafted policy limit comparison with state-of-the-art approaches. Overcooked-AI provides the benchmark context necessary for systematic evaluation, with established baselines and metrics enabling direct comparison of XAI approaches. Second, intrinsically explainable learned policies remain unexplored for real-time collaboration. Previous work used either Wizard-of-Oz methods without actual ML policies \cite{zhang_investigating_2023}, hand-crafted policies that don't effectively represent deployable systems \cite{paleja2021utility} or post-hoc explanation systems applied on opaque policies \cite{wang2024utility}. HA$^2$'s architecture bridges this gap, since its hierarchical structure enables explanation generation while its learned policies achieve competitive performance. Third, explanation modality effects in fast-paced environments remain largely unexplored. While some explanation systems have been evaluated separately on different collaborative environments, systematic comparisons between explanation modalities (without content differences) in benchmark environments remain absent from the literature. Real-time collaboration may benefit, for example, from audio delivery -- which avoids competing for visual attention during dynamic tasks, yet this has not been empirically evaluated.

By generating explanations directly from HA$^2$'s subtask selections and comparing text versus audio modalities across multiple sessions, we provide the first systematic evaluation of XAI-based support from state-of-the-art learned policies in an established collaborative benchmark.

%% file: chapters/methodology.tex
\section{Methodology}

This work investigates two primary research questions that address critical gaps in explainable AI for human-agent collaboration. First, we examine how XAI-based support generated from intrinsic explainability impacts human-agent collaboration in fast-paced environments (\textbf{RQ1}), where the temporal demands of real-time coordination may fundamentally alter explanation utility compared to slower-paced tasks. Second, we investigate whether explanation modality (specifically text versus audio delivery) affects collaboration outcomes (\textbf{RQ2}), recognizing that modality choice may be particularly consequential in visually demanding tasks where text explanations compete for limited attentional resources.

Drawing on prior literature, we formulated three hypotheses to guide our investigation. We hypothesize that explanations will improve team performance compared to no explanations (\textbf{H1}), based on evidence that transparency facilitates coordination in collaborative contexts. We further hypothesize that audio explanations will reduce cognitive overhead compared to text in visually demanding tasks (\textbf{H2}), as audio delivery may allow participants to process agent intentions without shifting visual attention from the primary task. Finally, we hypothesize that explanations will improve subjective agent perception (\textbf{H3}), consistent with findings that explainable agents are rated more positively regardless of performance effects.

We generate real-time explanations from the Hierarchical Ad Hoc Agents (HA$^2$) architecture and evaluate their impact on human-agent collaboration in Overcooked-AI across text and audio modalities. This section establishes the theoretical foundations for explanation generation, describes our agent implementation, and details the experimental design.

\subsection{Theoretical Foundations}

Collaborative tasks in Overcooked-AI are formalized as Decentralized Partially Observable Markov Decision Processes (Dec-POMDPs), defined by the tuple $\{S, A, P, r, O, \Omega, N, \gamma\}$ where $S$ is the shared state space, $A = A_1 \times A_2 \times \cdots \times A_N$ is the joint action space, $P(s'|s,a)$ defines state transitions, $r(s,a)$ provides shared rewards (appropriate for fully cooperative settings), $O$ represents agent observation spaces, $\Omega(o|s',a)$ gives observation probabilities, $N$ is the number of agents, and $\gamma \in [0,1]$ is the discount factor. Partial observability arises primarily from decentralized decision-making: while agents observe similar environmental states, each must commit to actions without knowing their teammate's concurrent choice or internal decision-making process.

To enable explanation generation, we leverage hierarchical reinforcement learning policies that decompose agent behavior into interpretable levels. Our agent uses:

\begin{itemize}
    \item \textbf{Subtask Space} $Z = \{z_1, z_2, \ldots, z_K\}$: A predefined set of semantically meaningful goal states that correspond to human-interpretable collaborative objectives (e.g., ``pick up onion from dispenser'', ``place onion in pot'').
    
    \item \textbf{High-Level Policy} $\pi_H(z|o)$: O $\rightarrow$ Z: Selects appropriate subtasks given current observations, with subtask selections reflecting strategic decisions about task priorities and coordination.
    
    \item \textbf{Low-Level Policies} $\pi_W^i(a|o,z)$: O $\times$ Z $\rightarrow$ A: Execute primitive environment actions to complete assigned subtasks, translating high-level intentions into concrete behaviors.
\end{itemize}

This hierarchical decomposition enables natural explanation generation because subtask selections operate at a level of abstraction that aligns with human task understanding. Unlike primitive actions, which represent moment-to-moment decisions, subtasks capture meaningful intentions that may persist over multiple timesteps. When the Manager selects subtask $z_i$, this decision can be translated into human-understandable explanations (``collecting ingredients'', ``serving soup'') rather than streams of primitive action descriptions.

However, a critical constraint must be acknowledged: when hierarchical policies use model-free reinforcement learning, the agent lacks explicit world models or planning capabilities. The Manager policy selects subtasks \emph{reactively} based on current observations, not through deliberative planning. Consequently, subtask selections represent the agent's current high-level intention at the moment of explanation, not commitments to future behavior. The agent may change its subtask selection in subsequent timesteps as new observations arrive, without internal representations of why changes occurred or what alternatives were considered. Despite this limitation, hierarchical task decomposition allows us to provide explanations that reflect the agent's actual decision-making process while still making sense to human teammates without the necessity of post-hoc rationalizations.

\subsection{Agent Implementation}

The HA$^2$ framework \cite{arocaouellette2025implicitlyaligninghumansautonomous} implements the hierarchical structure described above through a Manager-Worker architecture. The Manager policy $\pi_H(z|o)$ selects from twelve domain-specific subtasks in Overcooked-AI: (1) pick up onion from dispenser, (2) pick up onion from counter, (3) pick up dish from dispenser, (4) pick up dish from counter, (5) place onion in pot, (6) place onion on counter, (7) get soup from pot, (8) place dish on counter, (9) get soup from counter, (10) place soup on counter, (11) serve soup, and (12) waiting/unknown.

We use the original HA$^2$ Worker policies provided by the authors as pre-trained PyTorch models. For the Manager, we trained a custom policy using Proximal Policy Optimization (PPO) and self-play specifically on the Counter-Circuit layout. During training, both agents received their environmental observation plus their teammate's current subtask choice, enabling the Manager to potentially adapt its decisions based on the teammate's high-level intention. The Manager policy runs every timestep, outputting subtask choices that define the Worker policy responsible for primitive action selection at each timestep.

\begin{figure}
    \centering
    \includegraphics[width=.8\linewidth]{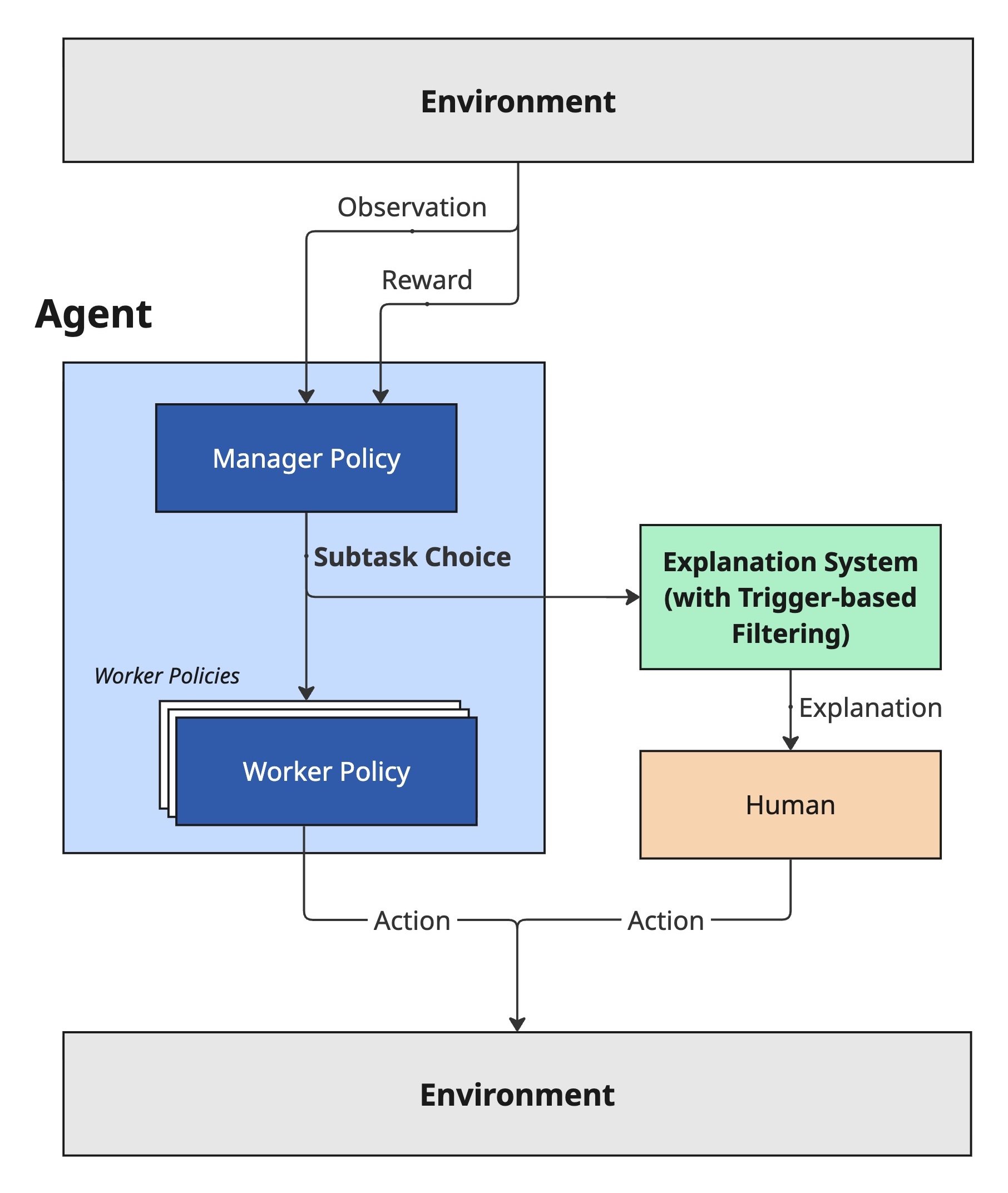}
    \caption{System architecture demonstrating how the Manager's subtask selections serve dual purposes in our XAI approach: directing the Worker's primitive actions and generating human-interpretable explanations. The trigger-based filtering system ensures explanations are delivered only at coordination-relevant moments rather than every timestep.}
    \label{fig:architecture}
    \Description{System architecture diagram showing the flow from environment to explanation generation. At the top, the Environment provides Observation and Reward to an Agent box. The Agent contains two components: a high-level Manager policy that selects subtasks, and a low-level Worker policy that outputs actions back to the Environment. The Manager's subtask output branches to both the Worker policy internally and to an external Explanation System box. The Explanation System processes the subtask information and generates explanations directed toward a Human figure. Both the Agent (via Worker) and Human provide Actions as input back to the Environment, completing the loop.}
\end{figure}

\subsection{Explanation Delivery System}

Converting Manager subtask selections into human-directed explanations required addressing temporal dynamics revealed through pilot testing with eight graduate students. As illustrated in Figure~\ref{fig:architecture}, the Manager's subtask outputs flow both to the Worker policy (for action execution) and to our explanation generation system (for human communication). Pilot testing revealed critical issues with continuous explanation display: first, explanations generated every timestep made it difficult for participants to understand them, let alone react accordingly. Second, minor policy fluctuations generated explanations with no coordination value, and participants reported focusing on gameplay rather than processing the constant stream of information. Continuously displaying Manager outputs also proved overwhelming, since the high-level policy runs every timestep and frequently changes selections even during stable task execution. While this behavior is natural for model-free policies, it proved to be confusing when communicated as human-like intentions. These findings motivated our trigger-based approach that generates explanations only at coordination-relevant moments.

We implemented a priority-based trigger system that generates explanations primarily in moments where revealing the agent's high-level intention is relevant to its teammate:

\begin{enumerate}
    \item \textbf{Blocking}: Triggers when the human occupies the agent's target tile or stands one tile away from its intended path, communicating navigation intent through templates like ``Moving around you to [subtask]'' or ``Excuse me, I want to [subtask].''
    
    \item \textbf{Critical Path}: Triggers when the agent selects subtasks essential for task progression (e.g., placing the final onion needed to complete a soup or collecting a ready dish from a pot), using templates such as ``I'll [subtask]'' to signal strategic coordination points.
    
    \item \textbf{Distance Threshold}: Triggers when the agent travels a predetermined number of tiles since the last explanation, providing status updates like ``I'm going to [subtask]'' during extended movement sequences.
    
    \item \textbf{Subtask Change}: Triggers when the Manager switches to a new subtask (excluding transient ``unknown'' states during policy transitions), announcing behavioral shifts with templates like ``Now I'll [subtask].''
\end{enumerate}

Each trigger maps to natural language templates that frame subtask descriptions appropriately for their coordination context (see Appendix A of supplementary materials for complete template specifications). A six-second cooldown prevents excessive communication regardless of trigger activation frequency. In the case of the text condition, each explanation remains visible for four seconds.

\textbf{Modality Implementation:} We deliver identical explanation content through two modalities. The \textbf{Text condition} displays explanations in a semi-transparent black bar at the bottom of the game interface, prefixed with ``Agent:'' in white text. The \textbf{Audio condition} converts identical text to speech using browser-native synthesis with automatic voice selection based on participant language preference (Portuguese or English) and speech rate optimized during pilot testing. Audio condition participants completed a validation task before gameplay: correctly entering a randomly generated number spoken by the text-to-speech system confirmed audio functionality. Text condition participants viewed a sample interface screenshot showing explanation placement.

\subsection{Experimental Design}

\begin{figure}
    \centering
    \includegraphics[width=1.0\linewidth]{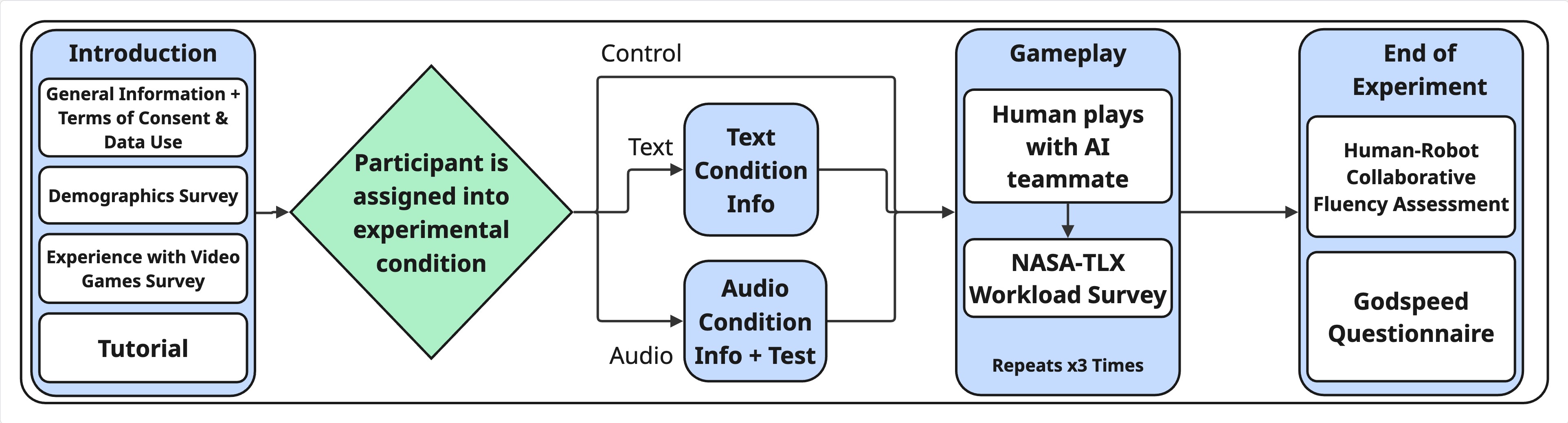}
    \caption{Experimental procedure flow showing blocked randomization for condition assignment. Participants complete introduction surveys before being randomly assigned to Control, Text, or Audio conditions, then play four gameplay sessions with intermediate assessments, followed by post-experiment questionnaires.}
    \label{fig:website}
    \Description{Flowchart diagram showing the three-phase experimental procedure. The first phase (Introduction) includes four sequential steps: General Information/Terms of Consent, Demographics Survey, Experience with Video Games Survey, and Tutorial with Game Controls and Text. After introduction, participants are assigned to experimental conditions via a decision diamond. The Control condition leads directly to Gameplay phase. Text and Audio conditions each lead to their respective "Condition Info" pages before Gameplay. The Gameplay phase shows participants playing with AI teammate, completing NASA-TLX Workload Survey after each session, repeated 3 times for 4 sessions total. The final phase (End of Experiment) includes three components: Human-Robot Collaborative Fluency Assessment, Godspeed Questionnaire, and Completion Screen.}
\end{figure}

Our between-subjects design assigned participants to three conditions via blocked randomization (block size 6, two participants per condition per block): \textbf{Control} (no explanations), \textbf{Text}, and \textbf{Audio}. Each participant completed four 80-second gameplay sessions in Overcooked-AI's Counter-Circuit layout at 10 FPS. We selected Counter-Circuit for three reasons: (1) coordination complexity -- the central counter obstacle requires continuous movement coordination and creates natural blocking scenarios where explanations could aid navigation; (2) strategic depth -- optimal performance requires dividing labor around the counter, enabling assessment of whether explanations facilitate role negotiation; (3) experimental control -- restricting to one layout eliminates layout-specific effects as a confounding variable for this initial study on the impact of XAI support. 

After each session, participants completed the NASA-TLX workload assessment \cite{hart1988development}. Following all gameplay, they evaluated the agent using the Godspeed Questionnaire \cite{bartneck2009measurement} (measuring anthropomorphism, animacy, likeability, perceived intelligence, and perceived safety) and a deployed adaptation of the Human-Robot Fluency Assessment \cite{hoffman2019evaluating} (changes are detailed in Appendix~\ref{app:fluency-adaptation}). Instructional manipulation checks embedded in the first and third NASA-TLX surveys and both post-experiment questionnaires verified participant attention. Complete survey instruments are provided in Appendix B of the supplementary materials.

We developed a web-based platform integrating Overcooked-AI with our explanation delivery system. The platform consists of a Flask backend with Firebase Firestore for data persistence and a browser-based frontend rendering gameplay via Phaser.js. The system was deployed on a virtual machine, supporting up to 15 concurrent sessions for remote participation. We collected game scores, participant action counts, network latency metrics, and complete survey responses. Participants achieving less than 80\% accuracy on instructional manipulation checks, who experienced technical issues (extreme network latency) or who did not complete all parts of the experiment (gameplay sessions and surveys) were excluded from analysis.

\begin{figure}[t]
    \centering
    \includegraphics[width=1.0\linewidth]{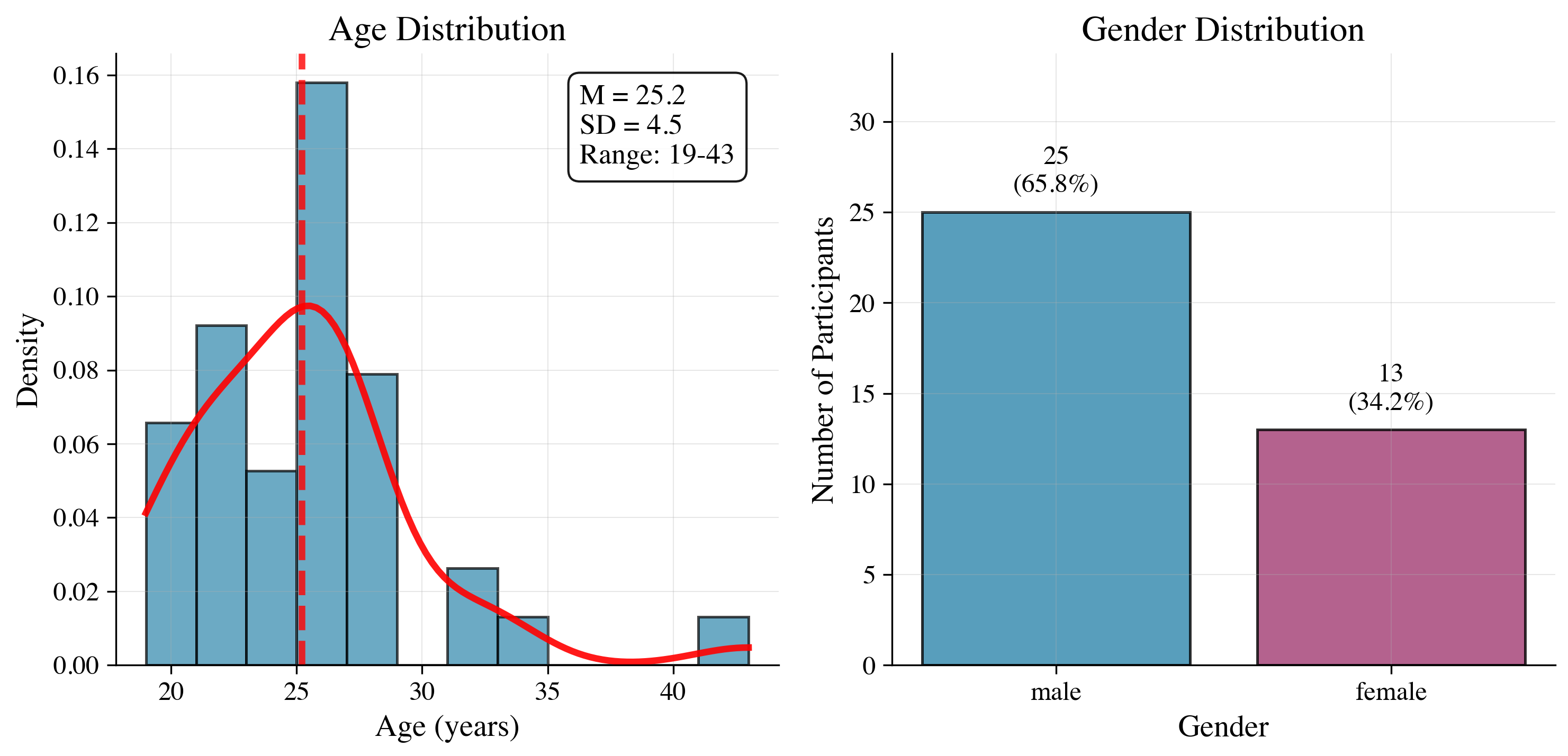}
    \caption{Participant demographics showing a young adult sample (M=25.2 years, SD=4.5) with male majority ($\approx$66\%). Age distribution spans 19-43 years with peak density around 25, while gender distribution shows 25 male and 13 female participants.}
    \label{fig:demographics}
    \Description{Two panel figure showing participant demographics. Left panel shows age distribution as a histogram with density across ages 20-40, peak density around age 25, mean 25.2, standard deviation 4.5, range 19-43 years. Right panel shows gender distribution as a bar chart with 25 male participants and 13 female participants, representing approximately 66 percent male and 33 percent female.}
\end{figure}

\subsection{Participants and Procedure}

\begin{figure}[t]
    \centering
    \includegraphics[width=0.75\linewidth]{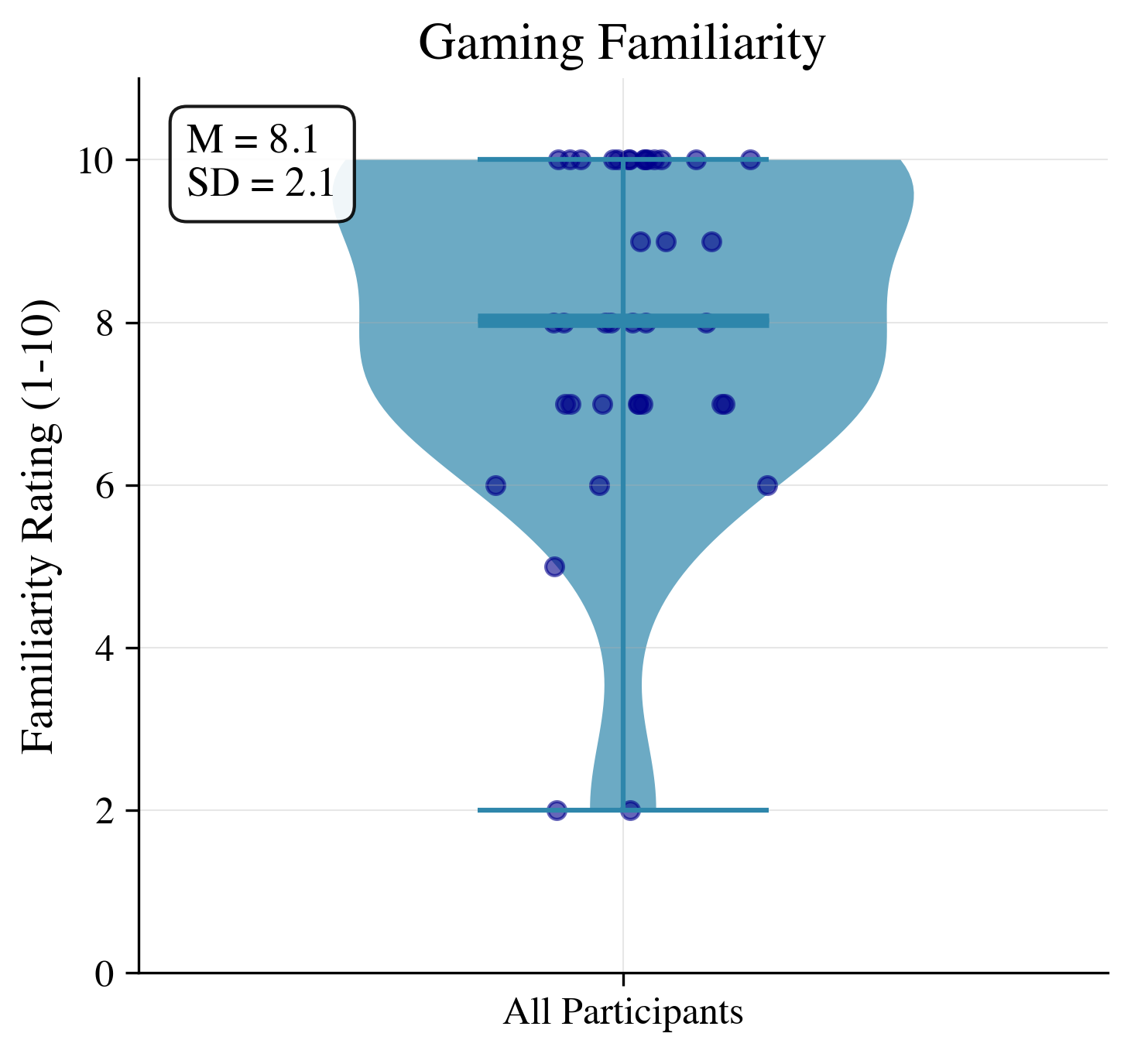}
    \caption{Gaming familiarity distribution demonstrating a gaming-expert sample. Participants reported high familiarity scores (M=8.1 on 1-10 scale, SD=2.1), with most clustered between 7-10, indicating substantial videogame experience that may influence explanation utility.}
    \label{fig:gaming}
    \Description{Violin plot showing distribution of self-reported gaming familiarity scores on a 1-10 scale. The distribution shows most participants clustered in the upper range between 7-10, with mean of 8.1 and standard deviation of 2.1. Individual data points are overlaid on the violin plot, showing concentration of responses at high familiarity levels, indicating a gaming-expert sample.}
\end{figure}

For our IRB-approved experimental study, we targeted a sample size of n=30 based on similar studies in human-agent collaboration research \cite{paleja2021utility} and practical recruitment constraints, expecting to detect large effects given the apparentness of the interventions. Anticipating potential connectivity issues during remote participation, we recruited 41 participants via convenience sampling within the local academic community (departmental emails and social media groups), yielding 38 valid participants after quality screening based on minimum accuracy on survey attention checks.

Post-hoc sensitivity analysis revealed our final sample provided 80\% power to detect large effects for H1 (Cohen's d$\geq$0.86), very large effects for H2 (d$\geq$1.12), and large effects for H3 (Cohen's f$\geq$0.53). Our observed effects were smaller -- H1: d=0.59 (medium), H2: d=0.14 (negligible) -- confirming we were underpowered for medium effects. We therefore frame this work as exploratory research establishing baseline effect sizes for future confirmatory studies.

Participants accessed a web-based platform, completing: (1) consent and demographics surveys, (2) videogame experience assessment, (3) tutorial video and text instructions, (4) condition assignment and modality-specific instructions, (5) four gameplay sessions with intermediate NASA-TLX assessments, and (6) post-experiment questionnaires. The complete experimental flow is shown in Figure~\ref{fig:website}.

Our final sample comprised 38 participants (ages 19-43, M=25.2, SD=4.5; approximately 66\% male; Control n=14, Text n=14, Audio n=10), as shown in Figure~\ref{fig:demographics}. Participants reported high gaming familiarity (M=8.1 on a 1-10 scale, SD=2.1), visualized in Figure~\ref{fig:gaming}, with 92\% self-reporting a score $\geq$ 6/10 and 84\% playing video games at least weekly. Unlike Paleja et al. \cite{paleja2021utility}, who used post-hoc performance clustering, we characterize participants based on this self-reported gaming familiarity, revealing a predominantly gaming-experienced sample. No significant baseline differences emerged across conditions in age, gaming familiarity, or gender distribution.

%% file: chapters/results.tex
\section{Results and Discussion}

We evaluated explanation effects on human-agent collaboration across three hypotheses: whether explanations improve performance (H1), whether modality matters (H2), and whether explanations improve subjective experience (H3). Our analysis reveals a nuanced picture: while explanations provided no immediate performance benefits for our gaming-expert participants, different modalities influenced adaptation patterns and produced a robust modality-specific cost in the felt working-alliance bond, even without any significant cognitive load differences -- illuminating future applications of XAI-based support for human-agent teaming.

\subsection{Performance Effects: Expertise and Adaptation}

An independent samples t-test comparing the control condition against combined explanation conditions showed a trend towards better performance of participants who did not receive any explanations (t(36)=-1.53, p=0.136, Cohen's d=-0.51), though this difference did not reach statistical significance. As shown in Table~\ref{tab:h1_performance}, the control condition scored higher (M=125.00, SD=18.19, n=14) than explanation conditions (M=114.17, SD=22.59, n=24), with a medium effect size suggesting practical relevance despite not reaching statistical significance.

\begin{table}[t]
\centering
\caption{Performance Measures: Control vs. Explanation Conditions (H1)}
\label{tab:h1_performance}
\setlength{\tabcolsep}{4pt}
\small
\begin{tabular}{@{} l c c c c c @{}}
\toprule
\textbf{Measure} & \textbf{Control} & \textbf{Explanations} & \textbf{t} & \textbf{p} & \textbf{d} \\
\midrule
Game Score    & 125$\pm$18 & 114$\pm$23 & $-1.53$ & .136 & $-0.51$ \\
Learning Rate & 4.6$\pm$9.0 & 9.3$\pm$7.5 & $-1.76$ & .087 & 0.59 \\
\bottomrule
\end{tabular}
\\[6pt]
\begin{minipage}{\linewidth}
\footnotesize
\textit{Note}: Explanations = Combined explanation conditions (Text n=14, Audio n=10). Learning rate = linear slope across sessions. Values shown as M$\pm$SD.
\end{minipage}
\end{table}

The identified trend is directionally consistent with Paleja et al.'s finding that expert performance can degrade with XAI support \cite{paleja2021utility}, as experienced users may quickly develop efficient mental models that explanations may disrupt. Our experiment population, made up of participants predominantly experienced with videogames (M=8.1 familiarity on 1-10 scale, SD=2.1), may have identified the optimal Counter-Circuit strategy -- using the middle counter for rapid object transfers -- rendering status-based explanations redundant once coordination patterns stabilized. Critically, we found no significant cognitive load differences across conditions (see Table~\ref{tab:h2_workload}), suggesting the performance trend stems from explanation interference with fluid coordination rather than raw processing demands.

\begin{figure}[t]
\centering
\includegraphics[width=\linewidth]{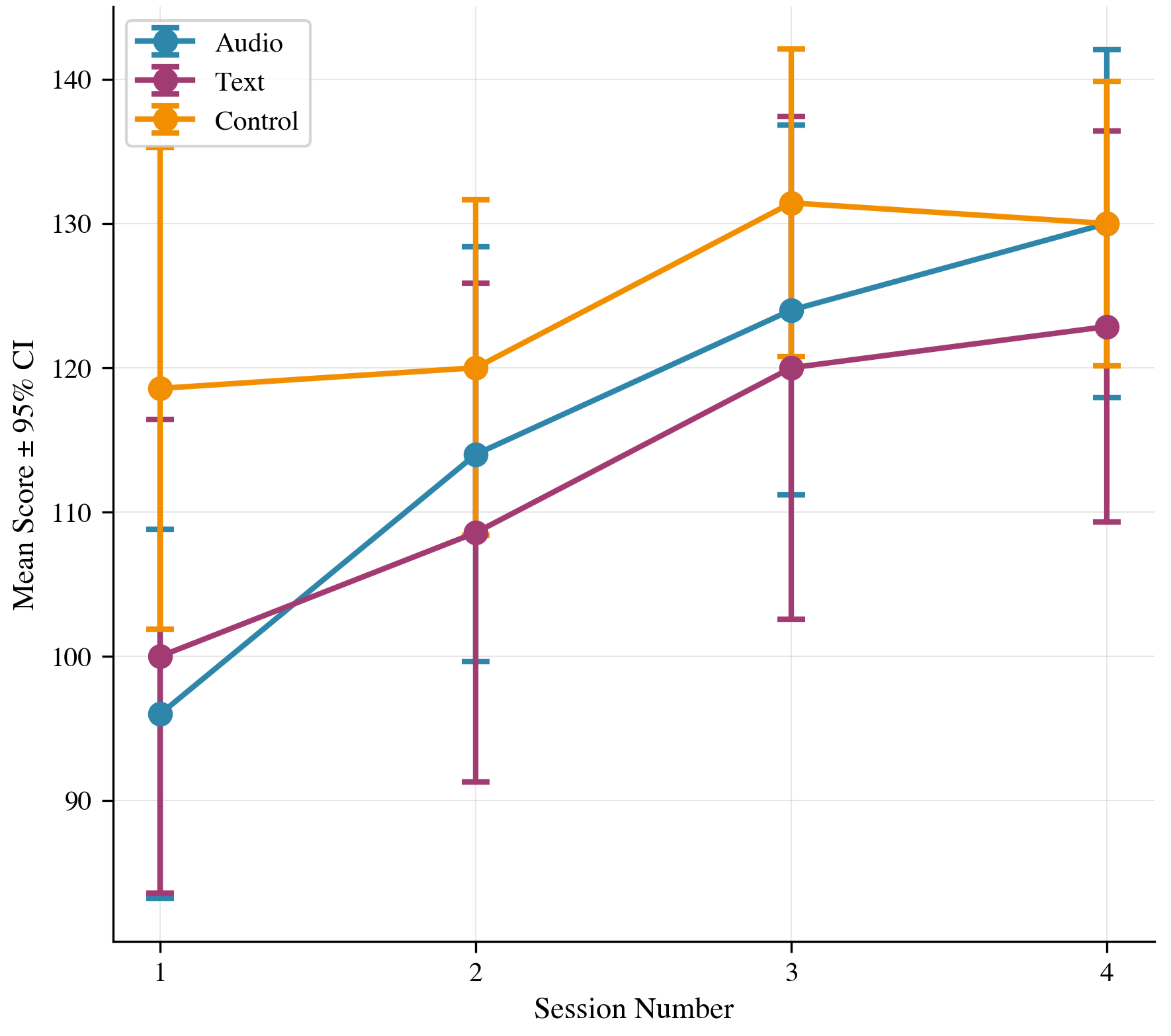}
\caption{Learning curves by condition showing equivalent peak performance but differential adaptation rates. Audio explanations produced the steepest learning trajectory despite no final performance advantage.}
\label{fig:learningcurves:condition}
\Description{Line graph showing learning curves across four gameplay sessions. The graph has three lines, one for each condition: Audio condition (blue) starts at 96 points in session 1 and reaches 130 points by session 4 with steepest slope; Text condition (orange) starts at 100 points and reaches 123 points with moderate slope; Control condition (green) starts at 118 points and reaches 130 points with shallowest slope. Error bars show standard deviations. Right panel  All conditions converge to similar performance by session 4 despite different learning rates.}
\end{figure}

However, examining improvement patterns reveals a more compelling story. Analysis of the scores over time (linear slopes across four sessions) revealed a clear monotonic pattern suggesting explanations accelerated participant adaptation: Audio participants had the fastest score improvement (M=11.20 points/session, SD=6.94), followed by Text (M=8.00, SD=7.85), then Control (M=4.57, SD=8.96). These results represented a 2.4$\times$ faster improvement rate for Audio versus Control. While our one-way ANOVA did not reach statistical significance (F(2,35)=2.000, p=0.151), the Control versus combined explanations comparison approached significance with a medium effect size (t(36)=-1.758, p=0.087, d=0.591). The Text versus Audio comparison also showed a small-to-medium effect (t(22)=-1.032, p=0.313, d=0.427), demonstrating systematic differences in adaptation speed despite equivalent peak performance. Figure \ref{fig:learningcurves:condition} illustrates these trajectories, with all conditions showing robust learning effects (r=0.335, p<0.001) but differing rates of improvement.

This pattern of equivalent peak performance alongside differential learning rates may suggest that explanations enhance human adaptation to agent behavior patterns rather than immediate task execution, though further research is needed to confirm this relationship. Expert teams may reach similar performance ceilings regardless of transparency, but explanations accelerate the mental model development that enables efficient coordination. The audio modality's showing of faster learning with equal peak scores to the Control condition supports our hypothesis that audio delivery might be promising for improved explanation delivery during fast-paced collaborative tasks.

\subsection{Modality Effects: Salience vs. Intrusiveness} \label{subsection:modality}

Comparing text versus audio explanations directly (H2) revealed equivalent performance (t(22)=0.329, p=0.745, d=0.136), with audio participants scoring marginally higher (M=116.00, SD=17.61) than text participants (M=112.86, SD=26.14). This negligible difference masks important salience effects: multiple text-condition participants reported not noticing explanations until later sessions, while audio explanations were universally perceived.

\begin{quote}
\textit{Participant 12 (Text): "I didn't notice the agent's explanations. I only realized the agent was communicating in the last session, and that was the session where I managed to score the most points. Since I was very visually focused on the middle of the screen, I ended up not seeing the messages below."}
\end{quote}

This feedback suggests text explanations could be easily disregarded, being treated as optional visual elements during fast-paced gameplay, while audio explanations demanded attention. This attentional difference may explain the modality's differential effects on learning rates and subjective measures (discussed below), even when both delivered identical content.

NASA-TLX analysis with Bonferroni correction ($\alpha$ = 0.0083) revealed no significant workload differences between modalities across six dimensions, as shown in Table~\ref{tab:h2_workload}. Mental demand was nearly identical (Text: M=3.12, SD=1.36; Audio: M=3.15, SD=1.34; t(22)=0.045, p=0.965, d=0.018). The largest observed difference was in effort ratings (Text: M=3.71, SD=1.13; Audio: M=3.23, SD=1.38; t(22)=-0.954, p=0.351, d=-0.395), which approached a small effect but remained non-significant after correction. 

These findings contradict concerns that audio explanations would impose higher cognitive costs; instead, their greater salience may reduce processing effort by making agent intentions immediately available without requiring visual attention shifts as with text. This higher salience becomes a helpful reasoning in light of the different subjective experience Audio participants reported compared to those in the other two experiment conditions, described below. 

\begin{table}[t]
\centering
\caption{Cognitive Workload: Text vs. Audio Explanations (H2)}
\label{tab:h2_workload}
\begin{tabular}{@{} l c c c c c @{}}
\toprule
\textbf{NASA-TLX} & \textbf{Text} & \textbf{Audio} & \textbf{t} & \textbf{p} & \textbf{d} \\
\midrule
Mental Demand & 3.1±1.4 & 3.2±1.3 & 0.05 & .965 & 0.02 \\
Effort & 3.7±1.1 & 3.2±1.4 & -0.95 & .351 & -0.40 \\
\bottomrule
\end{tabular}
\\[6pt]
\begin{minipage}{\linewidth}
\small
\textit{Note}: No significant differences after Bonferroni correction ($\alpha$=0.0083). Values on 1-7 scale, M±SD.
\end{minipage}
\end{table}

\subsection{Subjective Experience: Modality Matters for Felt Partnership} \label{subsection:bond}

One-way ANOVAs across the 13 Godspeed and Human-Agent Fluency subscales (H3), with Bonferroni correction ($\alpha = 0.0038$), revealed one effect that survived correction and one theoretically important trend. Both concentrate in the audio condition rather than spanning explanation modalities uniformly.

\begin{figure}[t]
\centering
\includegraphics[width=\linewidth]{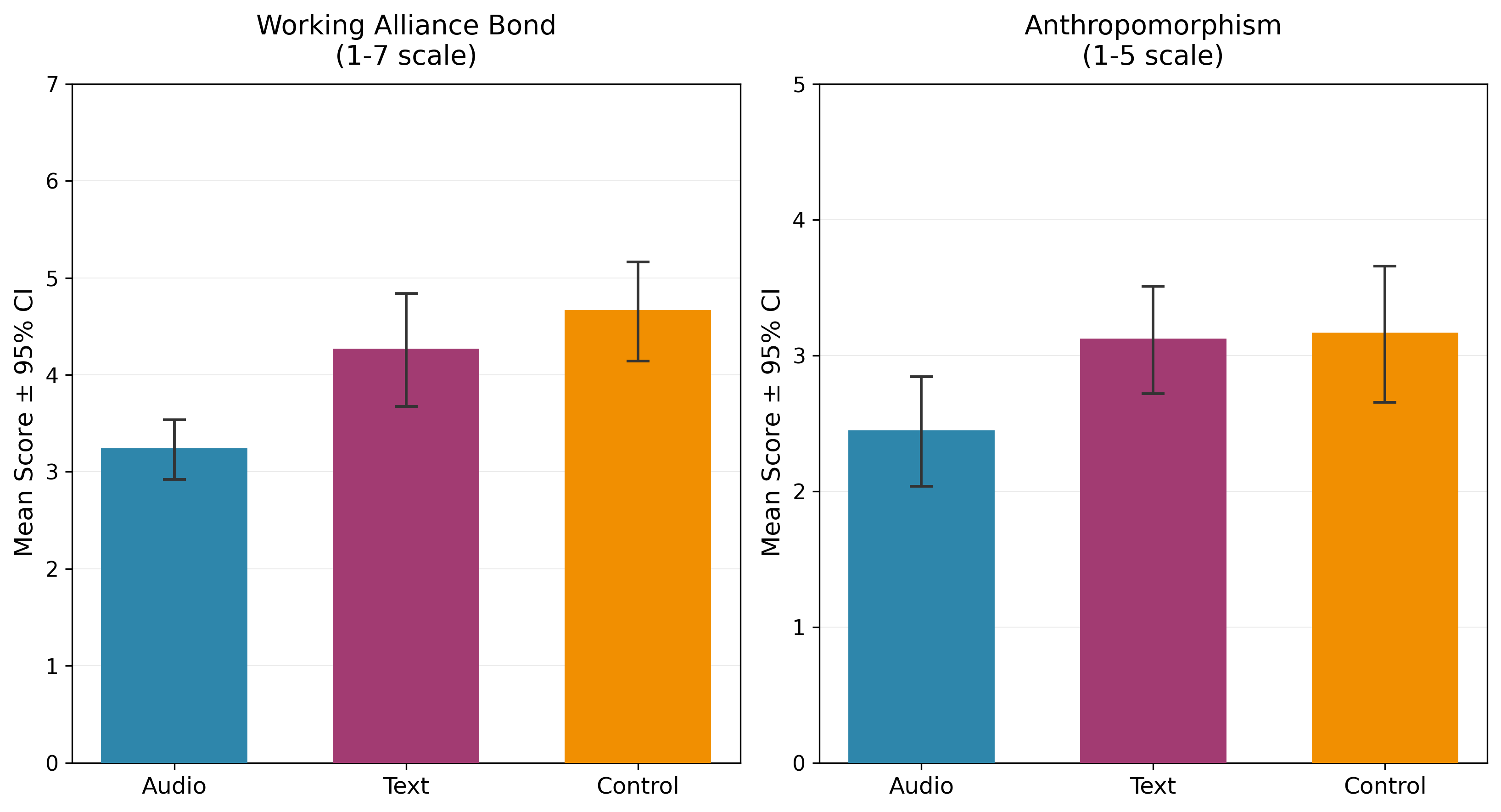}
\caption{Audio participants had a worse perception of their working alliance with the agent and also considered it the least anthropomorphic out of all conditions. Bond is on the 1--7 Fluency scale; Anthropomorphism on the 1--5 Godspeed scale.}
\label{fig:subjective}
\end{figure}

The Fluency Working-Alliance Bond subscale showed a robust condition effect (F(2, 35) = 6.98, p = 0.003, $\eta^2 = 0.285$; Cronbach's $\alpha = 0.76$). Audio participants rated the bond markedly lower (M = 3.23, SD = 0.50) than text (M = 4.26, SD = 1.11) or control (M = 4.65, SD = 0.97) on the 1--7 scale, as visualized in Figure~\ref{fig:subjective}. Variance was substantially lower in the audio cell than in the other two conditions (variance ratio = 5.01), so we additionally ran Welch's ANOVA, which does not assume homogeneity of variance and is preferred when this assumption is violated \cite{delacre2017psychologists}; the Welch result confirmed and strengthened the original finding (F(2, 22.86) = 12.61, p < 0.001).

Games-Howell pairwise post-hocs reveal that the effect is specific to the audio modality, rather than an effect of explanations: audio differs significantly from both control (M$_{\text{diff}} = -1.42$, p < 0.001) and text (M$_{\text{diff}} = -1.03$, p = 0.017), while control and text are statistically indistinguishable (M$_{\text{diff}} = +0.40$, p = 0.578). Text explanations carry no measurable working-alliance cost; the bond reduction is uniquely associated with spoken delivery.

The Godspeed Anthropomorphism subscale revealed a trend in the same audio-driven direction (F(2, 35) = 2.70, p = 0.081, $\eta^2 = 0.134$, Cronbach's $\alpha = 0.78$). Audio participants rated the agent as least anthropomorphic (M = 2.44, SD = 0.65), with text (M = 3.11, SD = 0.76) and control (M = 3.16, SD = 0.96) again nearly identical; the same asymmetric structure as the Bond finding, though without Bonferroni-level support.\footnote{Welch's ANOVA on Anthropomorphism yielded F(2, 22.78) = 3.44, p = 0.050, directionally consistent but still failing the Bonferroni threshold of $\alpha = 0.0038$.} The two subscales converge on the same conclusion: audio explanations, not explanations as a class, drive these subjective effects. 

\textbf{Mechanism.} The convergence of these two findings on the audio condition, combined with the absence of any cognitive-load difference between modalities (Section~\ref{subsection:modality}), rules out cognitive-intrusion explanations. Even so, audio explanations uniquely reduced the felt working-alliance bond (an effect counter to H3). While we cannot rule out that the absence of a bond effect under text partly reflects the lower perceived salience (as documented in Participant 12's feedback), we hypothesize the mechanism is social-relational, rather than attentional.

We propose that audio explanations promote the agent from a ``machine that helps'' to a (claimed) ``communicative partner'' in a way that text and silence (in the Control condition) do not. Text explanations can be ignored, with multiple participants in the text condition reporting not noticing them until later sessions (Section~\ref{subsection:modality}). Meanwhile, an agent that does not communicate makes no special partnership claim. Audio, by contrast, cannot be ignored: it is continuously perceived and frames the agent as actively communicating intent. 

This is analogous to human-human collaboration: if a person is calling out their teammate's attention and constantly communicating their intentions, these are perceived as faithful statements about future behaviour that should presumably be taken into account. The HA$^2$ Manager, however, is model-free and selects subtasks reactively at each timestep; it cannot maintain the consistent, committal coordination that this framing implies. The gap between ``agent speaking as a teammate'' and ``agent whose coordination is governed by reactive policy outputs'' registers, we argue, as a working-alliance violation. The Anthropomorphism trend in the same direction is consistent with this account: verbalization makes the claim-versus-reality gap more salient, lowering perceived human-likeness.

Participants in the audio condition described this gap explicitly:

\begin{quote}
\textit{Participant 20 (Audio): ``Across all 4 runs I scored 120 points, which frustrated me, since I imagined that over time the AI and I would become more synchronized. Unfortunately, this was not the case; I felt I was being blocked by the AI more with each successive run.''}
\end{quote}

This expectation of growing synchronization -- a hallmark of coordinative partnership -- is met with growing felt antagonism because of what was identified as a lack of commitment (or confusion) on the agent's part. This contrasts with the control condition, where participants engaged with a silent agent, who then appeared as a competent (even if quiet) collaborator, exemplified by feedback as below:

\begin{quote}
\textit{Participant 3 (Control): ``All in all, we work} [sic] \textit{well together and I think we became friends.''}
\end{quote}

The architectural source of this gap was, nevertheless, present in all conditions, and identifiable even in the text condition by participants who attended closely:

\begin{quote}
\textit{Participant 2 (Text): ``In the last two sessions, the agent kept saying it was going to do A and then did B or C, something I didn't see in the first two sessions. This caught my attention and made me pay more attention to what it was actually going to do, instead of what it said it was going to do.''}
\end{quote}

Nevertheless, it was probably much harder to miss the gap in the audio condition, since every stated intention that the reactive policy abandoned mid-execution was verbalized and consequently visibly contradicted by the agent's next action. The bond rupture is, on this account, the affective consequence of an unmissable claim-versus-reality gap inherent to model-free policy execution.

\textbf{Subscale reliability.} Three of the 13 subscales tested showed reliability concerns that warrant acknowledgement, none of which affects the headline finding. Godspeed Perceived Safety ($\alpha = -0.08$) is driven by the quiescent--surprised item, a documented weak link in Godspeed-V \cite{bartneck2009measurement}; this subscale was additionally tested with Welch's ANOVA due to variance heterogeneity, and remained non-significant in either case (Welch's F(2, 16.68) = 1.47, p = 0.258). Fluency Trust in Robot ($\alpha = 0.41$) is mechanically suppressed by its two-item structure rather than reflecting unreliable measurement; the result was non-significant (F(2, 35) = 0.25, p = 0.780). Fluency Positive Teammate Traits ($\alpha = 0.59$) reflects Hoffman's intentional dual-coding of Q10 across the Trust and Positive Traits subscales; also non-significant. The Working-Alliance Bond subscale that carries the headline finding is reliable ($\alpha = 0.76$).

\subsection{The Limits of Status-Based Explanations for Human-Agent Teaming}
\label{subsection:status}

By using the HA$^2$ agent's subtask choices as explanations of behaviour to a human teammate, we applied a type of XAI support characterized as ``status-based'' in past-work \cite{paleja2021utility}. Participant feedback revealed two gaps with this type of approach.

First, participants wanted explanations to carry predictive commitment beyond the immediate next subtask, as exemplified by the feedback below.

\begin{quote}
\textit{Participant 27 (Audio): ``The agent performed tasks in order to complete the team's goal while verbalizing its next action, so there is cooperation. However, it wasn't possible for me to actually coordinate my actions with the agent, since it could only guess what I wanted to do and I couldn't tell it to act differently. Ideal cooperation would be to divide tasks in a way that accelerates the process.''}
\end{quote}

Here, ``cooperation'' (which the agent is said to achieve) is distinguished from ``coordination'', which requires divisible roles, allocated tasks, and a stable expectation of who does what next. Status-based explanations communicate only the current subtask selection; they do not convey what the agent intends after that subtask completes, nor any commitment to maintain a role over the next several seconds of play. This connects directly to the claim-versus-reality gap discussed in Section~\ref{subsection:bond}: the HA$^2$ Manager re-samples its subtask at every timestep based on current state, so any verbalized intention reflects the policy's current output rather than a binding plan. Participants who heard ``I am going to do X'' under audio could not know that the very next timestep might select Y, and the violation of that implied commitment is what registered as bond rupture.

Second, participants wanted explanations to flow bidirectionally:

\begin{quote}
\textit{Participant 16 (Audio): ``I think the human-agent interaction would be better if the agent were} [sic] \textit{capable to notice a mistake by the human part and wait or suggest a fix action to the player. For example, the agent could say `Get out the way' instead of `I'm trying to serve a plate.'''}
\end{quote}

Rather than the agent narrating its own intent (``I'm trying to serve a plate''), they wanted the agent to model the human's situation and respond to it (``get out the way''). This is, in effect, a request for theory-of-mind. The HA$^2$ architecture, despite its hierarchical structure, contains no explicit model of the human teammate; the Manager's state observation includes the human's position but treats it as environmental geometry rather than as an agent with intentions to be reasoned about. Without a human model, the agent cannot acknowledge human errors, propose corrective actions, or negotiate role allocation -- which were precisely the affordances participants reached for when describing what would constitute ``ideal cooperation.''

Neither of these gaps is an artifact of the status-based paradigm per se; both are consequences of the underlying model-free, single-agent-perspective policy. A status-based explanation accurately reports what the policy is currently doing; if the policy is reactive and human-blind, the explanation will be reactive and human-blind too. The HA$^2$ approach demonstrates that hierarchical reinforcement learning can produce intrinsic explanations of agent behaviour without compromising task performance, which we consider a meaningful contribution. The present results suggest, however, that intrinsic explainability of a reactive policy is necessary but not sufficient for the explanation experience participants describe wanting from a teammate: predictive commitment requires a policy with a plan horizon longer than a single timestep, and bidirectional response requires a model of the human's actions and goals.

\textbf{Implications:} Taken together, these findings suggest that the most productive path forward is architectural rather than modal. The HA$^2$ approach demonstrates that hierarchical reinforcement learning can yield intrinsically explainable policies for collaborative tasks without sacrificing performance, and that even status-based explanations of this kind can accelerate human adaptation, particularly when delivered through audio. The modality-specific working-alliance cost we identified under audio explanations is not a reason to abandon spoken delivery, but a reason to make the policies that produce them more deserving of the partnership claim that spoken delivery implies. Two architectural changes could be promising in order to consider participant feedback: extending the Manager with a commitment horizon longer than a single timestep, so that verbalized intentions correspond to a binding plan rather than a transient sample; and incorporating an explicit model of the human teammate, so that explanations can flow bidirectionally and respond to human action rather than merely narrating the agent's own. Both extensions remain compatible with intrinsic explainability and represent, in our view, the most promising next step for status-based XAI in human-agent teaming.

%% file: chapters/conclusions.tex
\section{Conclusion}

This work establishes both the potential and constraints of leveraging intrinsically explainable machine learning architectures for collaborative AI systems. We provide the first systematic evaluation of XAI support generated from state-of-the-art learned policies in an established human-agent teaming benchmark, demonstrating that hierarchical reinforcement learning architectures can serve as both high-performing agents and sources of authentic explanations without sacrificing either capability. While performance and workload effects remained exploratory at our sample size (n=38), our subjective-experience analysis revealed a robust and Bonferroni-significant cost specific to spoken explanations: audio delivery reduced participants' felt working-alliance bond with the agent, a finding that constrains how and when status-based explanations should be deployed.

Our results reveal a nuanced picture of explanation utility among gaming experts. Simple status explanations from the HA$^2$ hierarchical policy showed trends toward lower immediate performance (d=-0.513, p=0.136), but with faster improvement rates (d=0.591, p=0.087). This pattern may suggest that XAI support enhances human adaptation to agent behaviour rather than immediate task execution, with transparency possibly accelerating mental-model development -- even though no significant cognitive workload difference was identified. Nevertheless, the lack of clear, direct improvement for ``experts'' receiving XAI-based support is aligned with past literature \cite{paleja2021utility}. 

The subjective-experience analysis, however, revealed that this salience advantage carried a specific relational cost. Participants in the audio condition rated their working-alliance bond with the agent markedly lower than those in either the text or control conditions (F(2,35)=6.98, p=0.003, $\eta^2$=0.285; Bonferroni-corrected $\alpha$=0.0038), with Games-Howell post-hocs confirming that the effect was specific to audio rather than to explanations as a class --- text bond ratings were statistically indistinguishable from control. A directionally consistent trend in Anthropomorphism ratings (p=0.081), also audio-driven, reinforces the asymmetry. 

We interpret this pattern as the affective consequence of an architectural mismatch: spoken explanations frame the agent as a communicative partner whose verbalized intentions warrant being treated as commitments, while the HA$^2$ Manager's reactive subtask re-selection cannot sustain that framing. The bond cost is therefore not an indictment of audio delivery as such, but of model-free policies whose actual coordinative behaviour falls short of what spoken partnership implies. The contrast with past literature showing that explanations improved participant perception of the agent \cite{paleja2021utility} can be explained by gap between the agent's communicated intentions and actual observed behaviour, something not explored before this work and which must be looked into in order to better determine the potential of applying XAI support for model-free RL as well as other  architectures not based on explicit planning. 

Our methodological contributions enable future XAI research in collaborative contexts. We demonstrate that: (1) intrinsically explainable architectures can generate authentic real-time explanations in benchmark environments without compromising task performance; (2) audio delivery ensures greater salience in fast-paced tasks without significant cognitive load cost; (3) trigger-based systems can manage communication frequency while maintaining naturalness; and (4) status-based explanations from reactive policies inherit the reactivity of the underlying architecture, imposing ceiling effects on what can be communicated coherently and on the relational quality the explanations can sustain. These insights suggest that effective collaborative XAI requires matching not only explanation sophistication to user expertise and task complexity, but also explanation richness to the underlying policy's actual capacity for plan commitment and partner modelling.

Future work should evaluate hierarchical explanations with novice users, in multiple environments, and using model-based architectures that can generate predictive rather than purely reactive explanations. Two architectural extension possibilities follow directly from participant feedback: a Manager with a commitment horizon longer than a single timestep, so that verbalized intentions correspond to a binding plan rather than a transient sample; and an explicit model of the human teammate, so that explanations can flow bidirectionally and respond to human action rather than merely narrate the agent's own. Longitudinal studies with more sessions could clarify whether learning-rate advantages eventually translate to sustained performance benefits, and whether the audio bond cost we observed attenuates as participants accumulate experience interpreting reactive agent speech.

Our work provides empirical evidence that hierarchical RL architectures offer a promising foundation for intrinsically explainable collaborative AI. The audio modality's twin signatures --- accelerated human adaptation alongside a measurable bond cost --- are not contradictory findings but a single architectural lesson: spoken explanations reach users effectively, but they only sustain partnership when the underlying policy can deliver what its delivery promises. This methodological contribution establishes a basis for systematically evaluating XAI approaches that meet both performance and partnership criteria in modern human-agent teaming applications.

\textbf{Limitations}: Our convenience sampling yielded a homogeneous sample of predominantly young, gaming-experienced participants from the academic community. While this enabled baseline findings suggesting XAI support does not universally improve collaboration among experienced users, generalizability to novice users, older adults, or populations with less gaming experience remains unknown. The sample's gaming expertise may have contributed to the observed trend that explanations did not improve immediate performance, as expert users may develop efficient coordination strategies that render simple status explanations redundant. The audio condition cell (n=10) was also smaller than the text and control cells (n=14 each) following our pre-registered IMC filter; while the Bond Alliance effect survived Welch's ANOVA, which does not assume equal variances or balanced cells, replication with balanced cells would strengthen confidence in the modality-specific interpretation. Audio explanations were delivered via browser-native text-to-speech, whose voice quality varies across browsers and operating systems and was not standardized in our deployment. Lower-quality or less natural-sounding speech may independently depress relational ratings, and we cannot fully separate this from the social-relational mechanism proposed in Section~\ref{subsection:bond}; replication with a controlled voice would help distinguish these accounts. Three of the thirteen subjective-experience subscales we analysed had reliability concerns documented in Section~\ref{subsection:bond}; none of them carried the headline finding, but their non-significant results should be treated as inconclusive rather than as evidence of absence. Finally, our sample size (n=38), while adequate for detecting large effects, was underpowered for the medium effects observed elsewhere in the analysis; broader replication is needed before drawing firm conclusions about XAI effectiveness in human-agent collaboration.